\documentclass[12pt]{article}
\pdfoutput=1
\usepackage{amsmath,amsfonts,amssymb}
\usepackage{hyperref}
\usepackage{graphicx}
\usepackage[font=small ,labelfont=bf]{caption}
\usepackage[font=footnotesize ,labelfont= small]{subcaption}
\usepackage{color}
\usepackage{epsfig}
\usepackage{psfrag}
\usepackage{slashed}
\usepackage{epstopdf}
\usepackage{ulem}
\usepackage{cite}
\usepackage{url}
\usepackage{braket}
\usepackage[lmargin=2cm,rmargin=2cm,bmargin=3cm,tmargin=2cm]{geometry}

\usepackage{color}
\newcommand \be{\begin{eqnarray}}
\newcommand \ee{\end{eqnarray}}

\makeatletter\@addtoreset{equation}{section}\makeatother

\DeclareMathOperator{\Tr}{Tr}

\DeclareMathOperator{\Ai}{Ai}

\def\bZ{\mathbb{Z}}

\newcommand{\beq}{\begin{equation}}
\newcommand{\eeq}{\end{equation}}
\newcommand{\bal}{\begin{equation}\begin{aligned}}
\newcommand{\eal}{\end{aligned}\end{equation}}

\newcommand{\eqn}[1]{(\ref{#1})}

\newcommand{\address}[1]{\vbox{\center\em#1}}
\renewcommand{\title}[1]{\vbox{\center\huge{#1}}\vspace{5mm}}

\newcommand{\cN}{{\mathcal N}}

\newcommand{\cO}{{\mathcal O}}

\begin{document}

\begin{titlepage}
\begin{center}
\phantom{}
%\vskip5mm 

\vspace{15mm}

\title{3d mirror symmetry as a canonical transformation}
\vspace{5mm}

\renewcommand{\thefootnote}{$\alph{footnote}$}

Nadav Drukker\footnote{\href{mailto:nadav.drukker@gmail.com}{\tt nadav.drukker@gmail.com}}
and Jan Felix\footnote{\href{mailto:jan.felix@kcl.ac.uk}{\tt jan.felix@kcl.ac.uk}}
\vskip 5mm 
\address{
Department of Mathematics, King's College London \\ 
The Strand, WC2R 2LS, London, UK}

\renewcommand{\thefootnote}{\arabic{footnote}}
\setcounter{footnote}{0}

\end{center}

\vskip5mm 

\abstract{
\normalsize 
\noindent 
We generalize the free Fermi-gas formulation of certain 3d $\cN=3$ supersymmetric 
Chern-Simons-matter theories by allowing Fayet-Iliopoulos couplings as well as mass 
terms for bifundamental matter fields. The resulting partition functions are given by simple 
modifications of the argument of the Airy function found previously. With these extra 
parameters it is easy to see that mirror-symmetry corresponds to linear canonical 
transformations on the phase space (or operator algebra) of the 1-dimensional fermions. 
}

\end{titlepage}

%\setcounter{tocdepth}{1}
%{\addtolength{\parskip}{-1.7ex}
%\tableofcontents 
%}

\section{Introduction and summary}
\label{sec:intro}

Supersymmetric localization leads to a dramatic simplification of the calculation of 
sphere partition functions (and some other observables) by reducing the infinite 
dimensional path integral to a finite dimensional matrix model \cite{Pestun2012,Kapustin2010} 
This matrix model can 
then be solved (sometimes) by a variety of old and new techniques to yield exact results. 
A particular application of this is to check dualities --- two theories which are equivalent 
(or flow to the same IR fixed point) should have the same partition function.

In practice, it is often very hard to solve the matrix models exactly, so dualities 
are checked by comparing the matrix models of the two theories and using integral 
identities to relate them. The first beautiful realization of this is in the 
Alday-Gaiotto-Tachikawa (AGT) correspondence, where the matrix models evaluating 
the partition functions of 4d $\cN=2$ theories were shown to be essentially identical to 
correlation functions of Liouville theory as expressed via the conformal bootstrap in 
a specific channel. $S$-duality in 4d was then related to the associativity of the OPE 
in Liouville, which is manifested by complicated integral identities for the fusion and 
braiding matrices \cite{Alday:2009aq, Dorn:1994xn, Zamolodchikov:1995aa, 
Ponsot:1999uf, Teschner:2001rv,Nekrasov:2002qd}.

Here we study 3d supersymmetric theories, which have several types of dualities, of which we 
will consider mirror symmetry and its $SL(2,\bZ)$ extension 
\cite{Intriligator1996, Hanany1997, Boer1997a, Boer1997, Witten2003, Assel2014}. 
Indeed one may use 
integral identities (in the simplest case just the Fourier transform of the 
sech function) \cite{Kapustin2010a} to show that the matrix models for certain mirror pairs are equivalent. 
But is there a way to simplify the calculation such that we can rely on a known 
duality of a model equivalent to the matrix model to get the answer without 
any work, as in the case of AGT?

Indeed for necklace quiver theories with at least $\cN=3$ supersymmetry (and one copy of 
each bifundamental field) there is a simple realization of the matrix model in terms of 
a gas of non-interacting fermions in 1d with a complicated 
Hamiltonian \cite{Marino2012}. 
The purpose of this note is to point out that the Hamiltonians of 
pairs of $\cN=4$ 
mirror theories are related 
by a linear canonical transformation.%
\footnote{In the specific case of ABJM theory, this was in fact already noted in \cite{Marino2012}, 
but here we prove it more generally.} 
Furthermore we show that the transformations 
between three known mirror theories close to $SL(2,\bZ)$, which is 
natural to identify with the $S$-duality group of type IIB, where the three theories 
have Hanany-Witten brane realizations.%
\footnote{We should mention of course also the 3d-3d relation \cite{Dimofte:2011ju}, 
which is closer in spirit to AGT and realizes mirror symmetry by 
geometrical surgery.}

In order to demonstrate this we generalize the Fermi-gas formalism of 
Mari\~no and Putrov to theories with nonzero Fayet-Iliopoulos (FI) parameters as 
well as mass terms for the bi-fundamental fields. This is presented in 
Section~\ref{sec:fermi} where we focus for simplicity on a two-node circular quiver.

In section~\ref{sec:mirror} we then present the action of mirror symmetry on the 
density operator of the Fermi-gas (the exponential of the Hamiltonian). We also 
outline the generalization to arbitrary circular quivers.
The generalization of this formalism to $D$-quivers and theories with symplectic 
gauge group will be presented in \cite{ADF}.

In the appendix we proceed to evaluate the partition function of the two-node quiver 
(and its mirrors). This was done for the theory without FI terms and bifundamental 
masses in \cite{Marino2012}, and we here verify that the calculation can be carried through 
also with these parameters turned on. The resulting expressions are not modified 
much and one can still express them in terms of an Airy function.

\section{Fermi-gas formalism with masses and FI-terms}
\label{sec:fermi}

In this section we review the Fermi-gas formulation \cite{Marino2012} 
of the matrix model of 3d supersymmetric field theories 
and generalize it to a particular $\mathcal{N}=4$
theory that includes all of the ingredients we will require for our study of mirror symmetry in the 
following section. This is a two node 
quiver guage theory with gauge group $U(N) \times U(N)$. Each node has a Chern Simons (CS) term with 
levels $k$ and $-k$. There is a single matter hypermultiplet transforming
in the fundamental representation of each $U(N)$ factor, and two matter hypermultiplets transforming in 
the bifundamental and anti-bifundamental representations of $U(N) \times U(N)$. 
The bifundamental fields have masses 
$m_1$ and $m_2$ and each node has a Fayet-Iliopoulos term with parameters 
$\zeta_1$ and $\zeta_2$. 

The matrix model for this theory is computed via localisation \cite{Kapustin2010}. The result can be 
easily derived by applying the rules presented for instance in 
\cite{Kapustin2010a,Gulotta2012,Yaakov2010}
\bal
\label{matrixmodel}
Z(N)= \frac{1}{(N!)^2}\int d^N \lambda^{(1)}d^N \lambda^{(2)}
&\frac{\prod_{i<j} 4\sinh^2 \pi(\lambda^{(1)}_i - \lambda^{(1)}_j)\, 
4\sinh^2 \pi(\lambda^{(2)}_i - \lambda^{(2)}_j)}
{\prod_{i, j} 2\cosh\pi(\lambda^{(1)}_i - \lambda^{(2)}_j + m_1) 
\,2\cosh\pi(\lambda^{(2)}_i - \lambda_j^{(1)}+ m_2)}
\\&\hskip1in
{}\times\prod_{i=1}^N 
\frac{ e^{2 \pi i \zeta_1 \lambda^{(1)}_i+\pi i k (\lambda^{(1) }_i)^2}
e^{2 \pi i \zeta_2 \lambda^{(2)}_i-\pi i k (\lambda^{(2) }_i)^2}}
{2\cosh \pi{\lambda^{(1)}_i} \, 2\cosh \pi{\lambda^{(2)}_i}}
\,.
\eal
The crucial step in rewriting this expression as a Fermi-gas partition function 
is the use of the Cauchy determinant identity
\beq
\frac{\prod_{i<j}(x_i -x_j)(y_i - y_j)}
{\prod_{i, j}(x_i - y_j)}
= \sum_{\sigma \in S_N}(-1)^{\sigma}\prod_{i=1}^N 
\frac{1}{(x_i - y_{\sigma(i)})}\,.
\eeq
Applying this to \eqn{matrixmodel} we may write the partition function as 
\bal
Z(N) = &\frac{1}{(N!)^2}\int d^N \lambda^{(1)}d^N \lambda^{(2)}
\sum_{\sigma_1 \in S_N}(-1)^{\sigma_1}
\prod_{i=1}^N \frac{1}{2\cosh \pi (\lambda^{(1)}_i - \lambda^{(2)}_{\sigma_1(i)}+ m_1)}
\\ &\times \sum_{\sigma_2 \in S_N}(-1)^{\sigma_2}
\prod_{i=1}^N \frac{1}{2\cosh \pi (\lambda^{(2)}_i - \lambda^{(1)}_{\sigma_2(i)}+ m_2)}
\prod_{i=1}^N \frac{ e^{2 \pi i \zeta_1 \lambda^{(1)}_i+\pi ik (\lambda^{(1) }_i)^2}
e^{2 \pi i \zeta_2 \lambda^{(2)}_i - \pi ik(\lambda^{(2) }_i)^2}}
{2\cosh \pi{\lambda^{(1)}_i}2\cosh \pi{\lambda^{(2)}_i}}\,.
\eal
A relabelling of eigenvalues 
$\lambda^{(2)}_i \rightarrow \lambda^{(2)}_{\sigma^{\smash{-1}}_1 (i)}$ 
allows us to resolve one of the sums over permutations, pulling out an overall factor of $N!$ giving
\bal
\label{Zisrho}
Z(N) &= \frac{1}{N!}\sum_{\sigma \in S_N}(-1)^\sigma \int d^N \lambda^{(1)}_i \, d^N \lambda^{(2)}_i 
\prod_{i=1}^N 
\frac{e^{2 \pi i \zeta_1 \lambda^{(1)}_i}e^{ i \pi k (\lambda^{(1)}_i)^2}}{2\cosh \pi{\lambda^{(1)}_i}}
\frac{1}{2\cosh\pi(\lambda^{(1)}_i - \lambda^{(2)}_i + m_1)}
\\&\hskip2in{}\times
\frac{e^{2\pi i \zeta_2 \lambda^{(2)}_i}e^{-i \pi k (\lambda^{(2)}_i)^2}}{2\cosh \pi{\lambda^{(2)}_i}}
\frac{1}{2\cosh\pi(\lambda^{(2)}_i - \lambda^{(1)}_{\sigma(i)}+ m_2)}
\\&
= \frac{1}{N!}\sum_{\sigma \in S_N}(-1)^\sigma 
\int d^N \lambda^{(1)}_i \, K\big(\lambda^{(1)}_i, \lambda^{(1)}_{\sigma(i)}\big).
\eal
Here we expressed the interaction between the eigenvalues $\lambda^{(1)}_i$ in terms 
of the kernel $K$, which can be considered the matrix 
element of the density operator $\hat K$ defined by

\beq
\label{K}
K(q_1, q_2) = \bra{q_1}\hat K \ket{q_2},
\qquad
\hat K = \frac{e^{2 \pi i \zeta_1 \hat q+\pi i k \hat q^2}}{2\cosh \pi \hat q}\,
\frac{e^{2 \pi i m_1 \hat p}}{2\cosh \pi \hat p}\,
\frac{e^{2 \pi i \zeta_2 \hat q- \pi ik \hat q^2}}{2\cosh \pi \hat q}\,
\frac{e^{2 \pi i m_2 \hat p}}{2\cosh \pi \hat p}
\,,
\eeq
where $\hat p$ and $\hat q$ are canonical conjugate variables 
$[\hat q,\hat p]=i\hbar$ with $\hbar=1/2\pi$ 
and we have made use of the elementary identities
\begin{align}
f(\hat q) \ket{q}&= f(q) \ket{q} \label{id1}
\\ 
e^{-2 \pi i m \hat p} f(\hat q) e^{2 \pi i m\hat p}&= f(\hat q - m) \label{id2}
\\ 
\bra{q_1}\frac{1}{\cosh \pi\hat p}\ket{q_2}&= \frac{1}{\cosh \pi(q_1 - q_2)}\,. \label{id4} 
\end{align}

To study the system in a semiclassical expansion it is useful to represent the operators in 
Wigner's phase space, where the Wigner transform of an operator $\hat A$ is defined as
\beq
\label{wigtrans}
A_W(q,p) = \int d q^\prime 
\Bra{q - \frac{q^\prime}{2}}\hat A \Ket{q + \frac{q^\prime}{2}}e^{i p q^\prime / \hbar}\,.
\eeq
Some important properties are 
\beq
\label{wigident}
(\hat A \hat B)_W = A_W \star B_W, \qquad \star 
= \exp \left[ \frac{i \hbar}{2}\left(\overleftarrow{\partial_q}\overrightarrow{\partial_p}
- \overrightarrow{\partial_p}\overleftarrow{\partial_q}\right) \right],
\qquad
\Tr(\hat A) = \int \frac{d q d p}{2 \pi \hbar}A_W \,.
\eeq
For a detailed 
discussion of the phase space approach to Fermi-gasses see \cite{Marino2012} and the 
original paper \cite{Grammaticos1979}. For a more general review of Wigner's phase space 
and also many original papers see \cite{Zachos2005}. 

In the language of phase space the kernel $\hat K$ \eqn{K} becomes
\beq
\label{KW}
K_W
= \frac{e^{2 \pi i \zeta_1 q+\pi i k q^2}}{2\cosh \pi q}
\star\frac{e^{2 \pi i m_1 p}}{2\cosh \pi p}
\star\frac{e^{2 \pi i \zeta_2 q- \pi ik q^2}}{2\cosh \pi q}
\star\frac{e^{2 \pi i m_2 p}}{2\cosh \pi p}\,.
\eeq
Clearly the partition function can be determined from the spectrum of $\hat K$ or $K_W$. 
The leading classical part comes from replacing the star product with a regular product. In 
the appendix we outline the calculation of the partition function, extending \cite{Marino2012}.

\section{Mirror symmetry}
\label{sec:mirror}

\begin{figure}[t]
\centering
\includegraphics[width=\textwidth]{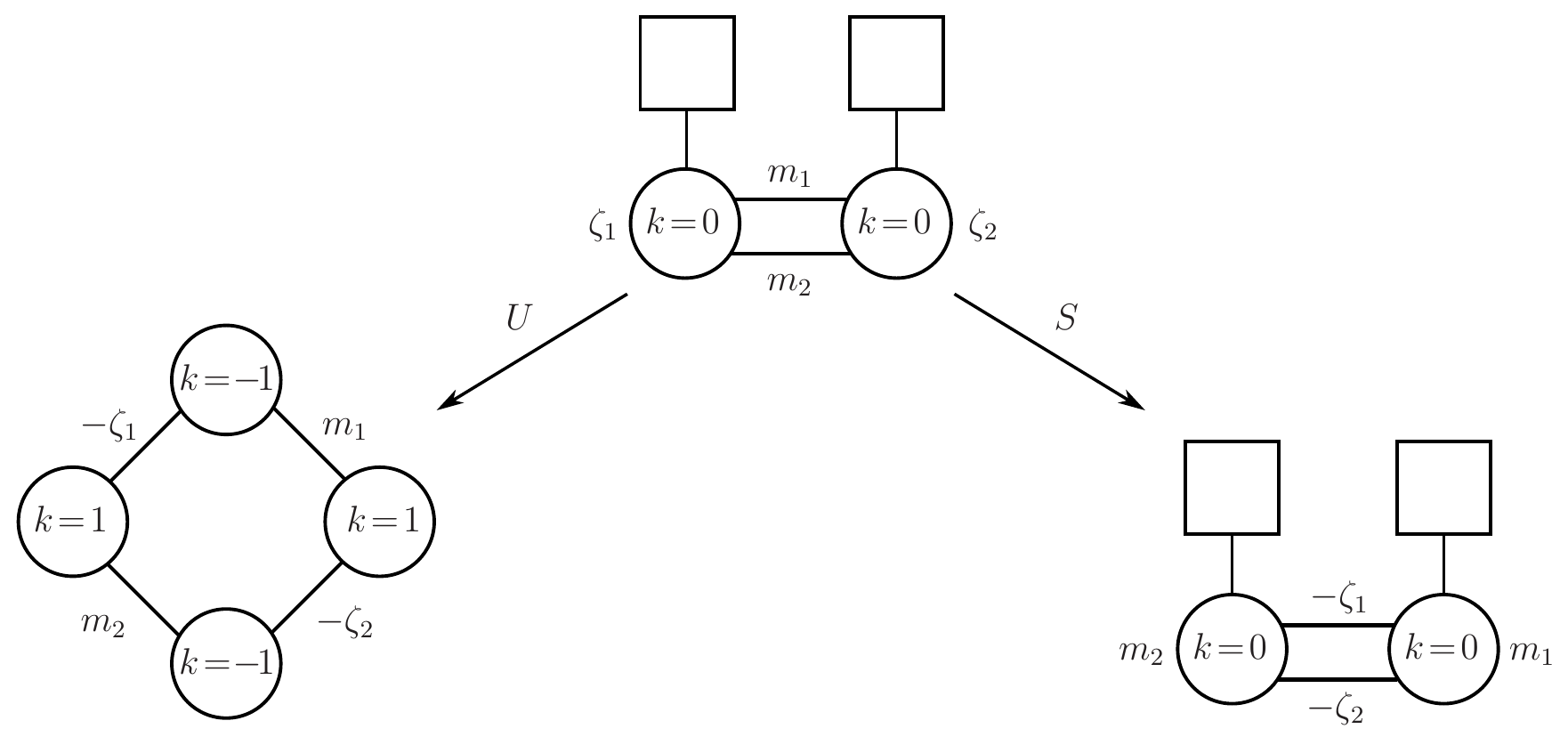}
\caption{\label{fig:mirror}%
Quiver diagrams summarising the two node theory we discuss in the text and its two mirror duals. 
Each circle represents a $U(N)$ vector multiplet, labelled inside by the CS level $k$ 
and outside by the FI parameter. Edges represent hypermultiplets. Those connecting two circles 
are bifundamental fields with the mass indicated next to them. The boxes represent $U(1)$ 
flavor symmetries of fundamental hypers, which in our examples are massless.}
\end{figure}

In this section we examine the theory studied in the previous 
section, with vanishing CS levels (see top quiver in Figure~\ref{fig:mirror}), 
where the density function \eqn{KW} becomes
\beq
\label{rhonoCS}
K_W = \frac{e^{2 \pi i \zeta_1 q}}{2\cosh \pi q}\star
\frac{e^{2 \pi i m_1 p}}{2\cosh \pi p}\star
\frac{e^{2 \pi i \zeta_2 q}}{2\cosh \pi q}\star
\frac{e^{2 \pi i m_2 p}}{2\cosh \pi p}\,.
\eeq
It has been known for a long time that this theory has two 
mirror theories, related in the IIB brane construction by $SL(2, \bZ)$ 
transformations \cite{Boer1997}. As we show, the density functions of these 
theories are simply related by linear canonical transformations.

\subsection{$S$ transformation}
\label{sec:S}

The first of the known mirror theories is one with identical matter content but with mass 
and FI parameters exchanged \cite{Boer1997a}
\beq
m_1 \rightarrow\tilde m_1= -\zeta_1\,, 
\qquad 
m_2 \rightarrow\tilde m_2=-\zeta_2\,, 
\qquad 
\zeta_1 \rightarrow \tilde\zeta_1= m_2 \,,
\qquad
\zeta_2 \rightarrow \tilde\zeta_2= m_1 \,.
\eeq
This is illustrated by the bottom right quiver in Figure~\ref{fig:mirror}.

At the level of the density function, this gives
\beq
\label{Srho}
K_W^{(S)}=
\frac{e^{2 \pi i m_2 q}}{2\cosh \pi q}\star
\frac{e^{-2 \pi i \zeta_1 p}}{2\cosh \pi p}\star
\frac{e^{2 \pi i m_1 q}}{2\cosh \pi q}\star
\frac{e^{-2 \pi i \zeta_2 p}}{2\cosh \pi p} \sim
\frac{e^{-2 \pi i \zeta_1 p}}{2\cosh \pi p}\star
\frac{e^{2 \pi i m_1 q}}{2\cosh \pi q}\star
\frac{e^{-2 \pi i \zeta_2 p}}{2\cosh \pi p}\star
\frac{e^{2 \pi i m_2 q}}{2\cosh \pi q}\,,
\eeq
where the last relation represents equivalence under conjugating by 
$\frac{e^{2 \pi i m_2 q}}{2\cosh \pi q}$.
We find that this density is the same as \eqn{rhonoCS} under the replacement
\beq
p \to q \,,
\qquad
q \to - p\,.
\eeq

\subsection{$U$ transformation}
\label{sec:U}
To get the second mirror theory we apply to \eqn{rhonoCS} the replacement
\beq
\label{Utrans}
p \rightarrow p + q\,,
\qquad
q \rightarrow - p \,.
\eeq
The result is%
\footnote{Note that the definition of the star product \eqn{wigident} is invariant 
under linear canonical transformations, and in particular under \eqn{Utrans}.}%
\bal
\label{Urho}
K_W^{(U)} &= 
\frac{e^{- 2 \pi i \zeta_1 p}}{2\cosh \pi p}\star
\frac{e^{ 2 \pi i m_1 (p+q)}}{2\cosh \pi (p+q)}\star
\frac{e^{-2 \pi i \zeta_2 p}}{2\cosh \pi p}\star
\frac{e^{2 \pi i m_2 (p+q)}}{2\cosh \pi (p+q)}
\\ 
&= \frac{e^{- 2 \pi i \zeta_1 p}}{2\cosh \pi p}\star
e^{-i \pi {q}^2}\star
\frac{e^{ 2 \pi i m_1 p}}{2\cosh \pi p}\star
e^{i \pi {q}^2}\star
\frac{e^{-2 \pi i \zeta_2 p}}{2\cosh \pi p}\star
e^{- i \pi {q}^2}\star
\frac{e^{2 \pi i m_2 p}}{2\cosh \pi p}\star
e^{i \pi {q}^2}\,.
\eal
In the second line we have made use of the identity 
\beq
\label{cs-shift}
e^{-\pi i{q}^2}\star f(p) \star e^{\pi i{q}^2}= f(p + q) \,.
\eeq
To read off the corresponding quiver theory from \eqn{Urho}, each 
$e^{i \pi (k q^2 + 2 \zeta q)}$ term can be associated to 
a $U(N)$ node with CS level $k$ and FI parameter $\zeta$, while each
$\frac{e^{2\pi i m p}}{2\cosh \pi p}$ comes from a bifundamental hypermultiplet 
with mass $m$. 
The transformed density operator corresponds therefore to a circular 
quiver with four nodes that have alternating Chern-Simons 
levels $k = \pm 1$ and vanishing FI parameters. The bifundamental multiplets connecting adjacent nodes have masses $\{- \zeta_1, m_1,-\zeta_2,m_2 \}$, 
as in the bottom left diagram of Figure~\ref{fig:mirror}.

As we discuss in the appendix, the partition function can be expressed in terms of  
$\Tr \hat K^l$ \eqn{Zconj}, which in phase space is given by an integral over $p,q$ \eqn{wigident}. 
Since we showed that mirror symmetry can be viewed just as a linear canonical 
transformation, which is a change of variables with unit Jacobian, it is clear that 
mirror symmetry preserves the partition function.

\subsection{$SL(2,\bZ) $}
\label{sec:sl2}

It is easy to see that the transformations we used in the previous sections close onto 
$SL(2, \bZ)$.
Indeed, defining $T=SU$ we find the defining relations
\beq
S^2 = - I\,,
\qquad
(ST)^3 = I\,.
\eeq
More general $SL(2,\bZ)$ transformations will give density operators with terms of the form
\beq
\frac{1}{\cosh \pi (a p +  b q) } \, .
\eeq
The cases with $a=0$ and $b=1$ or $a=\pm1$ and $b\in\bZ$ have a natural interpretation 
as a contribution of a fundamental field, or 
as we have seen in \eqn{cs-shift}, from conjugating the usual $1/\cosh\pi p$ by CS terms. 
But these manipulations cannot undo expressions one 
finds from a general $SL(2,\bZ)$ transformation of $K_W$. 
In these more general cases, the transformed 
density operator can still be associated to a matrix model, but it cannot be derived from 
any known 3d lagrangian.

This is also manifested in the IIB brane realization, where any $SL(2,\bZ)$ transformation will lead to 
some configuration of $(p,q)$ branes. Most of those do not have a known Lagrangian 
description \cite{Gaiotto2009}, but one could associate to them a matrix model \cite{Assel2014}, 
which would indeed lead to the transformed density operator.

\subsection{Mirror symmetry for generic circular quiver}
\label{sec:general}

The manifestation of mirror symmetry as a canonical transformation 
naturally generalises to the entire family of $\cN=4 $ circular quivers with an arbitrary 
number of nodes. 
Applying the Fermi-gas formalism, it is easy to see that the density function 
for such a theory with $n$ nodes is given by%
\footnote{The $\prod_\star$ product is defined by ordered star multiplication.}
\beq
\label{rhogen}
K_W = \prod_{a=1}^n\raisebox{-1.2ex}{}_\star \frac{e^{2 \pi i \zeta_a q}}{\left(2\cosh\pi q \right)^{N_a}}\star
\frac{e^{2 \pi i m_a p}}{2\cosh \pi p}\,,
\eeq
where $\zeta_a $ denotes the FI parameter of the $a$\textsuperscript{th} node, 
$N_a$ denotes the number of 
fundamental matter fields attached to the $a$\textsuperscript{th} node and $m_a$ 
denotes the mass of the bifundamental field connecting the $a$\textsuperscript{th} and 
$(a+1)$\textsuperscript{th} nodes. 

We can now apply the $S$ and $U$ transformations of the previous section, and look to see if 
the resulting density functions can again be interpreted as coming from the mirror gauge 
theories. Applying the $S$ transformation we get
\beq
K_W^{(S)}
= \prod_{a=1}^n\raisebox{-1.2ex}{}_\star 
\frac{e^{-2 \pi i \zeta_a p}}{\left(2\cosh\pi p\right)^{N_a} }\star
\frac{e^{2 \pi i m_a q}}{2\cosh \pi q}\,.
\eeq
This density function is that of a circular quiver theory with $\sum_{a=1}^n N_a$ 
nodes and $n$ fundamental matter fields. The fundamentals are attached 
to nodes which have FI parameters $m_a$, and are 
separated by $N_a-1$ other nodes. The masses of the bifundamentals connecting 
them add up to $-\zeta_{a}$.%
\footnote{\label{freedom}At the level of the matrix model, this additional freedom to choose mass parameters 
in the mirror theory simply amounts to the freedom to make constant shifts in the integration variables.}

Applying the $U$ transformation we get
\beq
K_W^{(U)}
= \prod_{a=1}^n\raisebox{-1.2ex}{}_\star \frac{e^{- 2 \pi i \zeta_a p}}{\left(2\cosh\pi p\right)^{N_a}}\star
\frac{e^{ 2 \pi i m_a (p + q)}}{2\cosh \pi (p + q)}
= \prod_{a=1}^n\raisebox{-1.2ex}{}_\star \frac{e^{- 2 \pi i \zeta_a p}}{\left(2\cosh\pi p\right)^{N_a}}
\star e^{-\pi iq^2}\star\frac{e^{ 2 \pi i m_ap}}{2\cosh \pi p}\star e^{\pi iq^2}\,.
\eeq
The mirror theory can be readily read off from this density function as a circular quiver 
theory with $\sum_{a=1}^n N_a + n$ nodes and no fundamental matter. Each node 
has Chern-Simons level $k= +1, -1$ or $0$. Further details concerning the mass parameters 
and value of the Chern-Simons level at each node can be read off in much the same 
way as for the previous example. 

A further generalisation we have not yet considered is to turn on masses for the 
fundamental fields. This corresponds to 
replacing each of the $(2\cosh\pi q)^{-N_a}$ 
in \eqn{rhogen} with a product of $N_a$ terms with masses $\mu_i$
\bal
\label{fundmasses}
e^{2 \pi i \zeta_a q}\star
\prod_{i=1}^{N_a}\raisebox{-1.2ex}{}_\star\frac{1}{2\cosh \pi (q + \mu_i)} 
&= e^{2 \pi i \zeta_a q}\star 
\prod_{i=1}^{N_a}\raisebox{-1.2ex}{}_\star
\left(e^{2 \pi i \mu_i p}\star \frac{1}{2\cosh \pi q }\star e^{-2 \pi i \mu_i p}\right)\\
&\hskip-1.2in{}
=e^{-2 \pi i \zeta_a\mu_1}e^{ 2 \pi i \mu_1 p}\star
\frac{e^{2 \pi i \zeta_a q}}{2\cosh\pi q}\star
e^{-2 \pi i \mu_1 p}\star
\prod_{i=2}^{N_a}\raisebox{-1.2ex}{}_\star
\left(e^{ 2 \pi i \mu_i p}\star
\frac{1}{2\cosh \pi q}\star
e^{ -2 \pi i \mu_{i} p}\right).
\eal
Where in the second line we chose to associate the FI term to the 
first fundamental field, picking up an overall phase.\footnote{There is a freedom to distribute 
the FI terms arbitrarily among the fundamental fields, leading to a different phase in front, 
see also footnote~\ref{freedom}.} 

Once we apply $S$ or $U$ transformations to \eqn{fundmasses} it becomes clear that 
these mass terms become additional FI parameters, as is expected. 

\subsection*{Acknowlegments}
We are grateful to Benjamin Assel and Marcos Mari\~no for useful discussions. 
N.D. is grateful for the hospitality of APCTP, of Nordita, CERN (via the 
CERN-Korea Theory Collaboration) and the Simons Center for Geometry and Physics, Stony Brook University 
during the course of this work. 
The research of N.D. is underwritten by an STFC advanced fellowship. The 
CERN visit was funded by the National Research Foundation (Korea).
The research of J.F. is funded by an STFC studentship ST/K502066/1

\appendix 
\section{From density to Airy function}
\label{sec:airy}

In this appendix we outline the computation of the large $N$ partition function for a 
theory with mass and FI parameters. Since the calculation 
follows closely the method outlined in
\cite{Marino2012} we will be rather cursory and refer the reader to \cite{Marino2012} 
for more detail, highlighting the new features that appear due to the FI and mass parameters.

To evaluate \eqn{Zisrho}, one notices that it 
combines to give products of $Z_l = \Tr (K_W)^l$ where the values of $l$ 
depends only on the conjugacy class of $\sigma$. Instead of summing over all permutations we can 
sum over conjugacy classes which have $m_l$ cycles of length $l$ and the combinatorics 
give
\beq
\label{Zconj}
Z(N)={\sum_{\{m_l\}}}^\prime \prod_l \frac{(-1)^{(l-1) m_l}Z_l^{m_l}}{m_l ! l^{m_l}}\,,
\eeq
where the primed sum denotes the restriction to sets that 
satisfy $\sum_l l m_l = N$. 
Following the usual analysis from statistical mechanics \cite{Feynman1972} 
we consider the grand canonical partition function given by 
\beq
\label{ximu}
\Xi(\mu) = 1 + \sum_{N=1}^\infty Z(N) e^{\mu N}
= \text{Exp}\left[-\sum_{l=1}^\infty \frac{(-1)^lZ_l e^{\mu l}}{l}\right] .
\eeq

We consider the density function \eqn{rhonoCS}, and using \eqn{id2}, \eqn{wigident} rewrite it as%
\footnote{It is also possible to rearrange the expressions such that one $p$ is not shifted and the 
other shifted by $\zeta_2$ and one $q$ is not shifted and the other shifted by $m_1$. We choose this 
more symmetric expression for later convenience.}
\bal
\label{Kwig}
K_W
&=e^{\pi i (m_1 \zeta_2 - \frac{1}{2} \zeta_1 m_1 -\frac{1}{2} \zeta_2 m_2) } e^{ \pi i ((2\zeta_1 +\zeta_2) q +m_1 p ) }\star
\frac{1}{2\cosh \pi (q- \frac{m_1}{2} )}\star 
\frac{1}{2\cosh \pi (p +\frac{\zeta_2}{2})}
\\&\hskip1.55in{}
\star \frac{1}{2\cosh \pi (q+\frac{m_1}{2})}
\star \frac{1}{2\cosh \pi (p-\frac{\zeta_2}{2})}
\star e^{\pi i (\zeta_2 q +( m_1 + 2 m_2)p ) }\,.
\eal
In order to get a hermetian Hamiltonian below, we specialize to the case
\begin{align}
\label{tracelessFImass}
\zeta_1=-\zeta_2=\zeta\,,
\qquad
m_1=-m_2=m\,,
\end{align}
and conjugate
\beq
\label{simtrans}
K_W \rightarrow (2 \cosh \pi (q-\tfrac{m}{2}))^{\frac{1}{2}}
\star e^{-\pi i (m p + \zeta q)} \star K_W \star 
e^{\pi i ( \zeta q + m p)}\star \frac{1}{(2 \cosh \pi (q-\frac{m}{2}))^{\frac{1}{2}}} \,.
\eeq
This gives the kernel
\beq
\label{Kwig2}
{K_W} 
= \frac{e^{- 2\pi i\zeta m}}{(2\cosh \pi (q-\frac{m}{2}))^{\frac{1}{2}}}\star 
\frac{1}{2\cosh \pi (p-\frac{\zeta}{2})}\star \frac{1}{2\cosh \pi(q + \frac{m}{2})}
\star \frac{1}{2\cosh \pi(p+\frac{\zeta}{2})} \star \frac{1}{(2\cosh \pi(q-\frac{m}{2}))^{\frac{1}{2}}} \, .
\eeq
The phase in $K_W$ leads to an overall phase $( e^{- 2\pi i\zeta m})^N$ in 
front of the partition function and can be removed and reintroduced at the end \eqn{ZNmzeta}.

\begin{figure}[t]
\centering
\begin{subfigure}{0.45\textwidth}
\includegraphics[width=\textwidth]{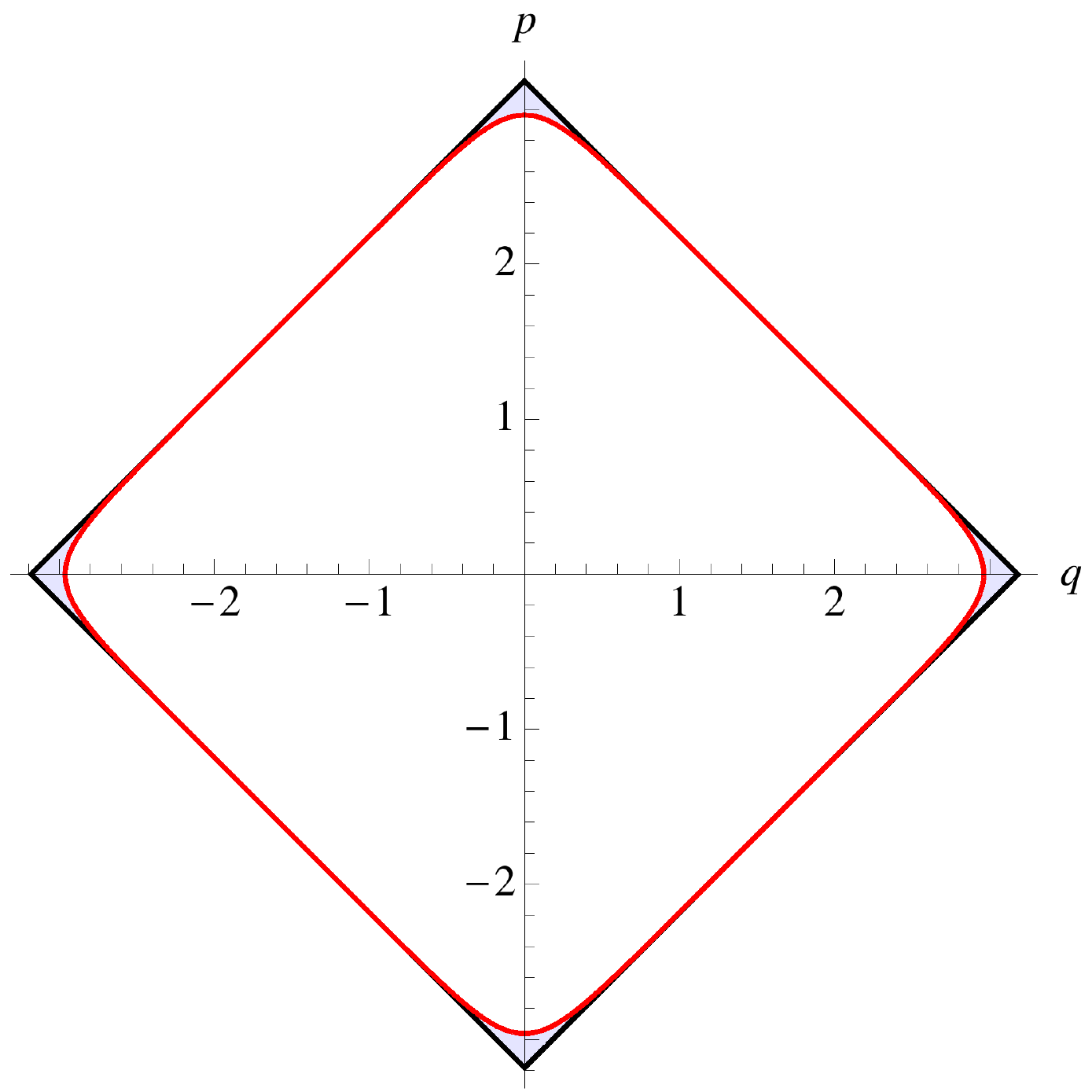}
\subcaption{\ \ \ }
\label{fig:massless}
\end{subfigure} \hspace{3mm}
\begin{subfigure}{0.45\textwidth}
\includegraphics[width=\textwidth]{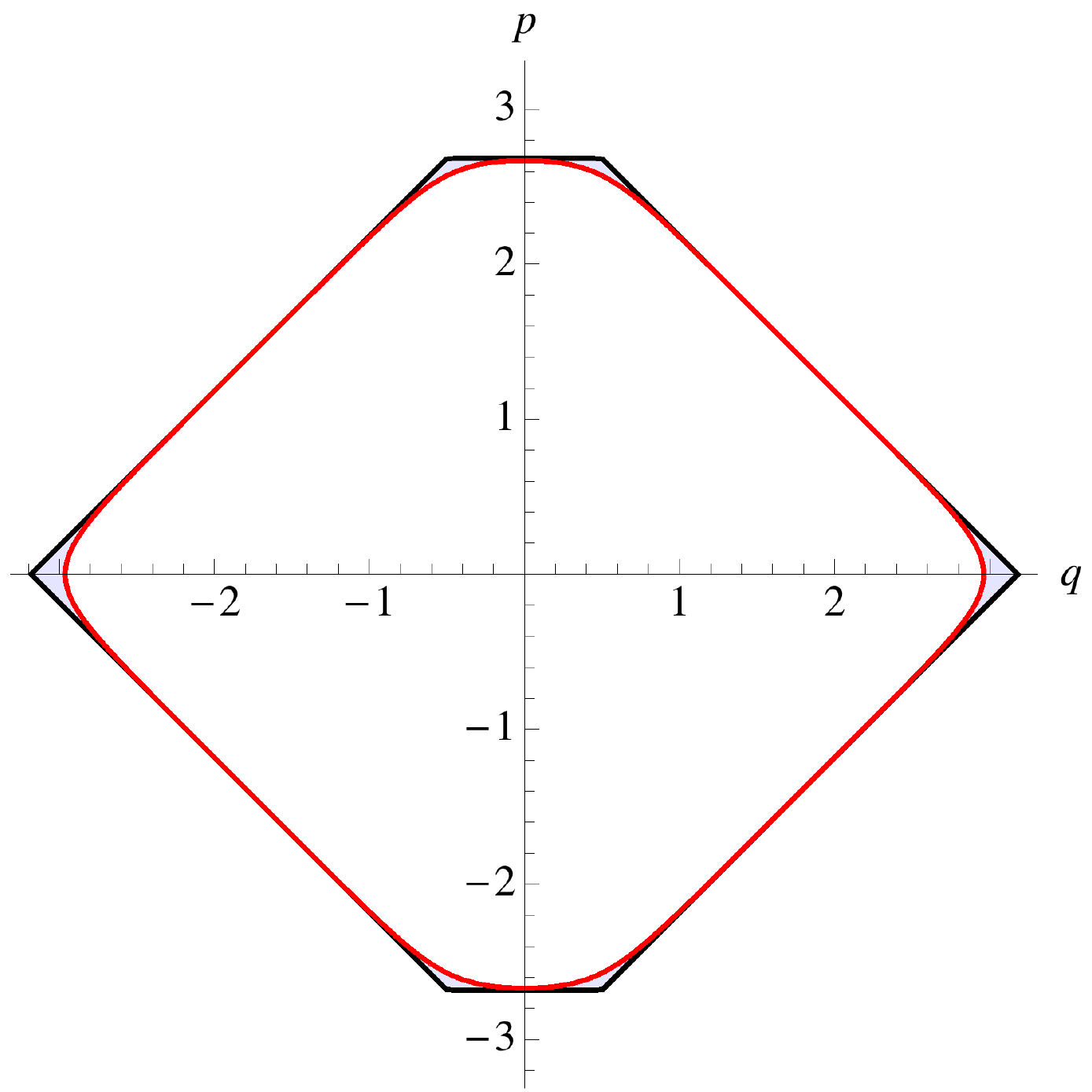}
\subcaption{\ \ \ }
\label{fig:massive}
\end{subfigure}
\caption{\label{fig:figure1}%
Fermi surfaces for $E=20$ and masses $m=0$ (a) and $m=1$ (b). 
The red lines are the exact classical Hamiltonians \eqn{Hexact} and black 
lines the polygon approximation \eqn{Happrox2}. 
The region between them is the leading 
order perturbative correction given in the first line of \eqn{vertexcorrections}.}
\end{figure}

Following Section~4 of \cite{Marino2012} we compute the partition function by studying the 
spectrum of the one particle Hamiltonian
\beq
\label{hamrho}
H_W = - \log_*K_W \,.
\eeq
To find an expression for $H_W$ one must perform a Baker-Campbell-Hausdorff (BCH) 
expansion of the logarithm in \eqn{hamrho}. Setting $\zeta=0$,%
\footnote{The effects of nonzero $\zeta$ 
and nonzero $m$ are completely analogous.}
the leading classical term in this expansion is simply
\beq
\label{Hexact}
H_{\text{cl}}= \log\left(2 \cosh \pi\left(q+\frac{m}{2}\right)\right) 
+ \log\left(2 \cosh \pi\left(q-\frac{m}{2}\right)\right) 
+2\log(2 \cosh \pi p)\, .
\eeq
For large $p,q$ this is
\beq
\label{Happrox2}
H_{\text{cl}} \approx \pi\left| q+\frac{m}{2}\right| + \pi\left| q-\frac{m}{2}\right| + 2\pi|p | \,.
\eeq
It is clear that the approximate Hamiltonian is independent of $m$ for $\left| q \right| > \frac{m}{2} $. 
In Figure~\ref{fig:figure1} we display the 
exact classical Fermi surface and polygonal approximation for a particular 
value of $E$ and with vanishing and non vanishing mass. The only change to the 
polygon from turning on the mass is the removal of the two triangles with 
$|p|>\frac{E}{2\pi}-\frac{m}{2}$
whose combined area is $m^2/2$. 

The number of states below the energy $E$ is given by the area enclosed by the curve 
$H=E$. Using the polynomial approximation this is just
\beq
\label{ne}
n(E) \approx \int dq dp \, 
\theta\!\left(E - \pi\left| q-\frac{m}{2} \right| - \pi\left| q-\frac{m}{2}\right|-2\pi|p| \right) = C E^2-\frac{m^2}{2},
\qquad
C= \frac{1}{2 \pi^2} \,.
\eeq
This expression is only approximate and gets corrected by accounting for the difference between 
\eqn{Happrox2} and the exact quantum Hamiltonian \eqn{hamrho}. 
We do this by modifying the number of states to
\bal
n(E) = C E^2 -\frac{m^2}{2}+ n_0 + n_\text{np}(E)\,, 
\eal
We outline the calculation of $n_0$ below. The main point is that it does not depend on 
$m$. $n_\text{np}$ denote nonperturbative, exponentially suppressed corrections at large 
$E$,%
\footnote{These also include the so called Wigner-Kirkwood corrections to the formula 
\eqn{ne}, which manifest as boundary integrals on the Fermi surface and are nonperturbative 
for the same reason as in \cite{Marino2012}}
but are such that $n(0)=0$. 
To satisfy this, $n_\text{np}$ clearly has to depend on 
$m$, but this will have no effect on our end result where we ignore the nonperturbative 
terms. 
The approximation \eqn{Happrox2} 
is valid where both $p$ or $q$ are large, and as shown in \cite{Marino2012}, 
the quantum corrections to \eqn{Hexact} (from the BCH expansion of \eqn{hamrho}) are also 
exponentially suppressed there. All the corrections are therefore associated 
with regions where either $p$ or $q$ are small, namely around the vertices of the polygon. 
We then consider the contributions to $n_0$ from each region separately, integrating in each case the 
(perturbative) corrections to the boundary. 
For instance, around $q\sim0$, $p\gg1$ the first quantum corrections of the Hamiltonian are given 
(up to terms exponentially suppressed in $p$) by
\beq
H_{\text{cl}} \to H_{\text{cl}} +\frac{\pi^2}{24}\left(\frac{1}{\cosh^2 \pi (q+\frac{m}{2})}-\frac{2}{\cosh^2 \pi (q-\frac{m}{2})}\right) \, .
\eeq
The difference in the area between the polygon and the quantum corrected Fermi surface 
for large $E$ approaches
\bal
\label{vertexcorrections}
\int_{-\infty}^\infty \frac{dq}{2\pi}
\Bigg[&\pi\left|q+ \frac{m}{2}\right| + \pi\left|q - \frac{m}{2}\right| 
- \log 2\cosh\pi\left(q+\frac{m}{2}\right)- \log 2\cosh\pi\left(q-\frac{m}{2}\right) \\
&{}-\frac{\pi^2}{24}\left(\frac{1}{\cosh^2 \pi (q+\frac{m}{2})}-\frac{2}{\cosh^2 \pi(q-\frac{m}{2})}\right)
\Bigg]
=-\frac{1}{24}\,.
\eal
As advertised, this is independent of $m$ (which can be seen by splitting the integral into two 
term with $q\pm m/2$ and shifting the integration variable).

The analog expression around the $p = 0$, $q \gg 1$ vertex is
\beq
\int_{-\infty}^\infty \frac{dp}{2\pi} \left(2\pi|p| - 2\log 2 \cosh \pi p - \frac{\pi^2}{48\cosh^2\pi p} \right) 
= -\frac{5}{48}\,.
\eeq
Summing over the contributions from all four regions we get
\beq
n_0 = 2\left(-\frac{1}{24} - \frac{5}{48}\right) = -\frac{7}{24}\, .
\eeq
It is not hard to see that all the higher order quantum corrections do not modify $n_0$ and 
in particular do not depend on $m$. See the discussion in Section~5.3 of \cite{Marino2012}.

From $n(E)$ it is easy to calculate the matrix model partition function. The grand canonical potential, the logarithm of \eqn{ximu} is
\beq
\label{jmu}
J(\mu) = \log \Xi(\mu) = \int_0^\infty dE \rho(E) \log(1 + e^{\mu-E})\,,
\qquad
\rho(E)=\frac{dn(E)}{dE}\,.
\eeq
At large $\mu$ this integral reduces to
\begin{align}
J(\mu) = \frac{C}{3} \mu^3 + \mu \left(n_0 - \frac{m^2}{2} + \frac{\pi^2 C}{3} \right) + 
A + \mathcal{O} (\mu e^{-\mu} ) \, .
\end{align}
$A$ is a constant that we will not concern ourselves with (it is studied in 
more detail in \cite{Marino2012} and subsequent papers). From this we can extract 
the canonical partition function%
\footnote{For a discussion of the integration contour, see \cite{Hatsuda2013}.}
\beq
Z(N) = \frac{1}{2 \pi i} \int_{- i \infty}^{i \infty} e^{J(\mu) - \mu N} 
= C^{-\frac{1}{3} } e^A \Ai\left[C^{-\frac{1}{3}}\left(N - n_0 - \frac{\pi^2 C}{3}+\frac{m^2}{2} \right)\right] 
+ \cO(e^{-N})\,.
\eeq
It is straight forward to include also the FI parameter $\zeta$. Remembering the extra phase in 
\eqn{Kwig2} one finds
\beq
\label{ZNmzeta}
Z(N) = C^{-\frac{1}{3}}e^{A-2 \pi i \zeta mN}
\Ai\left[C^{-\frac{1}{3}}\left(N - n_0 - \frac{\pi^2 C}{3}+\frac{m^2+\zeta^2}{2} \right)\right].
\eeq

One can treat quite general $\mathcal{N}=3$ necklace quiver theories in a similar fashion, 
subject to the technical constraint that the sum over FI parameters and the sum over bifundamental 
masses both vanish. The analog of \eqn{Happrox2} will again be a piecewise linear Hamiltonian
\begin{equation}
H\approx \sum_{i} | a_i q + b_i p + c_i |\,,
\end{equation}
The parameters $a_i$ and $b_i$ are determined by the 
Chern-Simons terms and $c_i$ are due to mass and FI terms. 
The volume of the corresponding polygonal Fermi surface is again of the form 
\begin{align}
C E^2 + B 
\end{align}
Similar arguments to those made above guarantee that the full $c_i$ 
dependence appears via a shift $B$ which can be found already in the polygonal approximation.

\bibliographystyle{utphys2}

\bibliography{References}

\end{document}